\documentclass[12pt,letterpaper]{article}
\usepackage[letterpaper]{geometry}
\usepackage{amssymb,amsmath}
\usepackage{mathtools}
\usepackage{amsthm}
\usepackage{bbm}
\usepackage{graphicx}
\usepackage{enumerate}
\usepackage{caption}
\usepackage{natbib}
\usepackage{url} 
\usepackage{bm}
\usepackage{color}

\begin{document}

\def\spacingset#1{\renewcommand{\baselinestretch}%
{#1}\small\normalsize} \spacingset{1}

\title{\bf Explainable Linear and Generalized 
  Linear Models by the Predictions Plot}       
\author{Peter J. Rousseeuw\\
  Section on Statistics and Data Science, 
  KU Leuven, Belgium}
\date{July 22, 2025} 
\maketitle

\begin{abstract}
Multiple linear regression is a basic 
statistical tool, yielding a prediction 
formula with the input variables,
slopes, and an intercept. But is it 
really easy to see which terms have
the largest effect, or to explain why the
prediction of a specific case is unusually
high or low? To assist with this the 
so-called predictions plot is proposed. 
Its simplicity makes it easy to interpret, 
and it combines much information. 
Its main benefit is that it helps 
explainability of the prediction formula as
it is, without depending on how the
formula was derived. The input variables 
can be numerical or categorical. 
Interaction terms are also handled, and the 
model can be linear or generalized linear.
Another display is proposed to visualize
correlations and covariances between 
prediction terms, in a way that is tailored 
for this setting.
\end{abstract}


\spacingset{1.45} 
\section{Introduction} \label{sec:intro}

Suppose someone has carried out a linear
regression and tells you they predict
the horsepower (\textit{hp}) of a car by 
the formula $$\widehat{hp} =
  2.466*\mbox{topspeed}
  - 13.13*\mbox{length}
  + 0.063*\mbox{displacement} -206.9$$
where \texttt{topspeed} is in miles per hour, 
\texttt{length} is in meters and engine 
\texttt{displacement} is in cubic centimeters.
(This example is worked out in the
Supplementary Material.)
Which of these input variables has the 
biggest effect on the prediction? It is not 
enough to look for the largest coefficient,
because these depend on the units of the
variables. To resolve this, the typical 
approach is to first standardize each input 
variable by dividing it by its standard 
deviation. That works fine, but cannot be 
done for categorical input variables. We 
would like to assess the effect magnitude 
of input variables visually, in a way that 
includes categorical input variables. That 
is the topic of the next section.

A closely related issue is the question: Why 
is the prediction for a particular case 
so high or so low? An answer may be required
when the prediction determines an important 
decision such as granting or denying a loan 
request, or treating a patient's medical 
condition by surgery or medication. 
The ability to answer such a
question is called explainability.
It is often stressed that a regression
tree is explainable because you can follow 
its branches, whereas most neural nets are 
not. Is a linear prediction explainable? 
On the one hand, a linear combination has 
a simple expression. On the other hand, how 
can you tell which of its terms were mainly
responsible? This is a topic that deserves
to be touched upon in a basic statistics 
class. We propose a visualization in 
Section~\ref{sec:casetoshow},
and look at correlations between
prediction terms in 
Section~\ref{sec:predscor}. We describe
some related methods in 
Section~\ref{sec:related}, and
Section~\ref{sec:conclusions} concludes.

In this note we are not concerned with the 
methodology used to obtain the fit.
We will consider the prediction formula as
a given, and study what it does. 
Our displays will not use the observed
response variable either, or describe how 
well the prediction approximates it. That
is addressed by other tools such as residual 
plots.

\section{Spread and orientation of 
         prediction terms} 
\label{sec:predictionsplot}

A linear prediction on numerical input 
variables $x_1,\ldots,x_p$ is of the form
\begin{equation}\label{eq:prednumeric}
  \mbox{total\_pred} = a + \sum_{j=1}^p b_j x_j
\end{equation}
where $a$ is the intercept and the $b_j$ are
slopes (coefficients). We assume that 
the input variables have already been 
transformed if needed, and that outlying 
cases have been deleted or corrected. 
Let us now center the terms with the $x_j$\,,
and denote them as $f_j := b_j(x_j - 
\overline{x}_j)$\,. The terms $f_j$ will be 
called {\it prediction terms} 
to distinguish them from the original 
input variables $x_j$\,. We can also 
compute the average total prediction 
$\overline{\mbox{total\_pred}}$ which is 
sometimes called the {\it centercept} 
\citep{Wainer2000}, a name coined by John 
Tukey, because for the {\bf center}ed 
input variables it is the 
inter{\bf cept}. Next we also center 
the total prediction and denote 
it as $f := \mbox{total\_pred} -
\overline{\mbox{total\_pred}}$\,
so that~\eqref{eq:prednumeric} becomes
\begin{equation}\label{eq:predlm}
  f = \sum_{j=1}^p f_j
\end{equation}
without any intercept. In 
fact~\eqref{eq:predlm} is more general
than~\eqref{eq:prednumeric} because a 
categorical variable also yields a 
prediction term $f_j$\,, 
that is obtained by coding its levels by 
binary dummies and adding up their 
predictions. We do not scale the 
prediction terms $f_1,\ldots,f_p$ 
so all terms are in the units of the 
total prediction. Finally, we measure the 
spread of each prediction term 
$f_j$ by its standard deviation.

Our first example uses the Top Gear data 
from the \textsf{R} package \texttt{robustHD} 
\citep{Alfons:robustHD}. It contains 
numerical and categorical variables about 
297 cars, scraped from the website of the 
British television show Top Gear.
We want to predict a car's fuel efficiency
from the time in seconds it takes to
accelerate from standstill to 60 miles per 
hour (\texttt{accel}), the variable 
\texttt{drive} which has three levels: 
rear-wheel drive, front-wheel drive, and
four-wheel drive (4WD), the car's 
\texttt{weight} in kilograms, and the type 
of \texttt{fuel} it uses (petrol or diesel).
Following \cite{Henderson1981} who analyzed
a similar dataset we predict gallons per 
mile (GPM), the inverse of miles per 
gallon (MPG). 

\begin{figure}[!ht]
\centering
\includegraphics[width=.75\textwidth]
   {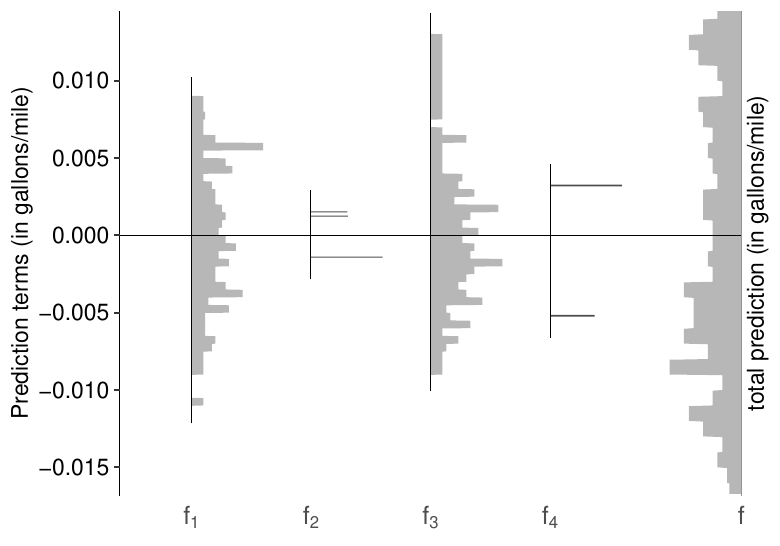}
\vspace{-2mm}
\caption{Top Gear data: an 
intermediate step toward visualizing the 
prediction term $f_1$ originating 
from the input variable \texttt{accel}, 
the term $f_2$ obtained from
\texttt{drive}, $f_3$ from \texttt{weight},
and $f_4$ from \texttt{fuel}.
On the right we see the total prediction 
$f$ of \texttt{GPM} computed as
in~\eqref{eq:predlm}.}
\label{fig:topgear_plain_predsplot}
\end{figure}

After a standard regression analysis,
briefly summarized in part A of the
Supplementary Material, we can compute  
the prediction terms $f_1$ 
obtained from \texttt{accel}, 
$f_2$ from \texttt{drive}, $f_3$ from 
\texttt{weight}, $f_4$ from \texttt{fuel}, 
and the total prediction $f$ which 
by~\eqref{eq:predlm} is the sum of the 
four terms. 
Figure~\ref{fig:topgear_plain_predsplot} 
is an intermediate step in the construction
of the proposed display. It plots $f_1$,
$f_2$, $f_3$, $f_4$ and $f$ next to 
each other. All the $f_j$ as well as 
$f$ are centered, and they all have the 
same units of gallons per mile.
The prediction terms $f_1$ and $f_3$ 
come from numerical input variables 
and take many values, so their 
distributions are shown as histograms.
The prediction terms $f_2$ and $f_4$ 
take very few distinct values, so they are
represented by bar charts. We clearly see
that the effect of prediction term 
$f_2$ is quite small compared to that of 
$f_1$. One might have expected that the 
average of prediction term 
$f_4$ which only takes two values would lie 
exactly in the middle of the two bars, but 
it doesn't because the bars have different
lengths, indicating that the higher 
prediction occurred more often than the 
lower one in this dataset.

\begin{figure}[!ht]
\centering
\includegraphics[width=.80\textwidth]
   {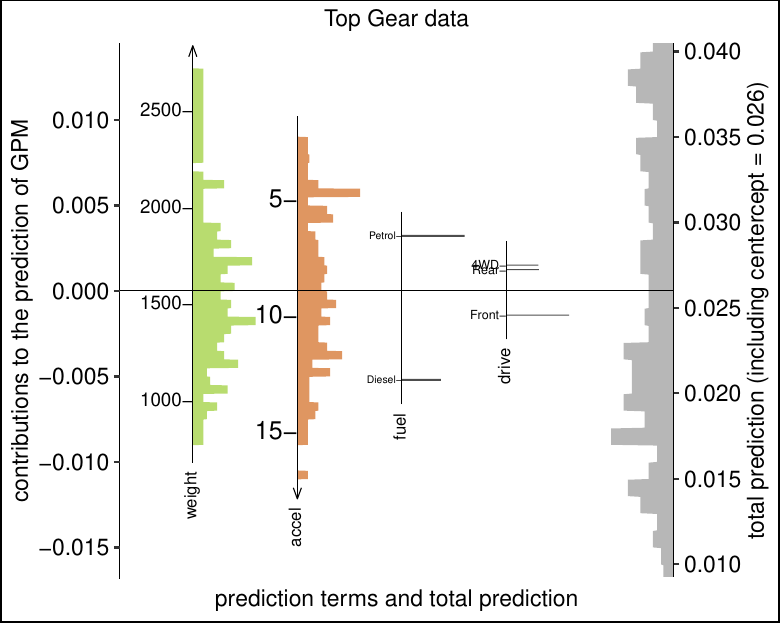}
\vspace{-2mm}
\caption{Predictions plot of the Top 
  Gear data. It shows the prediction terms
  originating from the input variables
  \texttt{weight}, \texttt{accel},
  \texttt{fuel} and \texttt{drive}.
  The prediction terms are centered, and
  ordered by decreasing standard deviation.
  They are all in the units of the vertical 
  axis on the left. The \texttt{weight}
  prediction term is shown in green with an
  upward arrow because its input variable
  has a positive coefficient, whereas 
  \texttt{accel} is brown with a downward
  arrow due to its negative coefficient.
  The categorical variables have no arrow.
  On the right we see the total prediction 
  in its units.}
\label{fig:topgear_predsplot}
\end{figure}

Figure~\ref{fig:topgear_plain_predsplot} 
already gives some intuition, but we can do 
more. It turns out that the \texttt{weight} 
prediction term has the largest 
standard deviation, followed by 
\texttt{accel}, \texttt{fuel}, and 
\texttt{drive}. In order to visualize this
we can order the prediction terms $f_j$ 
by decreasing standard deviation. 
This is an example of effect ordering 
for data displays \citep{Friendly2003}; 
see also \citep{Vanderplas2023} for another
instance of reordering variables.
Moreover, we can label the `anonymous' 
prediction terms in the scale 
of the original input variables, as 
in Figure~\ref{fig:topgear_predsplot}. 
For the numerical input variables 
\texttt{weight}
and \texttt{accel} we read off the values of 
$x_j$ without their slope coefficients.
Predictions from the categorical variables 
\texttt{fuel} and \texttt{drive} are labeled 
by their levels. The vertical axis on the 
left measures the contribution of each term 
$f_j$ to the total prediction. For
instance, increasing the weight of a car
from the average (1562 kg) to 
2000 kg, while keeping everything else 
fixed, increases the total
prediction by about 0.005 gallons/mile. 
Replacing a diesel engine by a petrol engine 
raises the prediction by 0.008
gallons/mile. The effects of four-wheel drive
and rear-wheel drive are about the same so
their labels overlap, whereas front-wheel 
drive decreases the prediction. 

Note that the \texttt{weight} line has an
upward arrow and its histogram is shown in
green, which indicates that increasing
\texttt{weight} yields a higher predicted
fuel consumption
(again, if the other characteristics 
remain the same). On the other hand, longer 
acceleration times yield lower 
predictions, as reflected
by the downward arrow and brown color.
The lines of the categorical variables 
\texttt{fuel} and \texttt{drive} have no
arrow because their levels are not 
ordered, and their distributions are
shown in neutral grey. The same 
will happen for
predictions from logical (Yes/No) 
variables and interaction terms.
The rightmost line with the total
prediction $f$ is labeled in its 
original scale that includes the 
centercept (which is
about 0.026 here). Its range is a bit 
truncated in the plot, to put the focus
on the prediction terms $f_j$\,. 
So we do not see its entire grey 
histogram, but that option is available.

We call Figure~\ref{fig:topgear_predsplot} 
a {\bf predictions plot}. The 
plural in ``predictions'' indicates that 
several predictions are plotted together.
The colors of the prediction terms 
can of course be changed by the user.
An advantage of the predictions 
plot over a list of standard
deviations of input variables 
is that it visualizes the relative effects 
of the input variables. 
We would also notice immediately if a 
variable were highly skewed, 
and any outlier would stick out like a 
sore thumb.

The predictions plot also works for 
generalized linear models, where the 
prediction~\eqref{eq:predlm} becomes
\begin{equation}\label{eq:predglm}
  f = g^{-1}\Big(\sum_{j=1}^p f_j\Big).
\end{equation}
Here $g$ is a monotone {\it link function}.
In logistic regression the link function is 
the logit function $g(p) = \log(p/(1-p))$, 
so~\eqref{eq:predglm} predicts a value 
between 0 and 1 using 
$g^{-1}(y) = 1/(1+\exp(-y))$.

To illustrate this we analyze the well-known 
Titanic dataset which contains information 
on the 1309 passengers of the RMS Titanic. 
For its history and visualizations see
\cite{Friendly2019}.
The data are freely available from 
\url{https://www.kaggle.com/c/titanic/data}
and the \textsf{R} package \texttt{classmap}
\citep{classmap}.
The binary response variable indicates whether
the passenger survived or was a casualty. 
We want to predict survival by a logistic
regression on the variables \texttt{pclass}, 
\texttt{sex}, \texttt{age}, \texttt{sibsp}, 
and \texttt{parch}.  
Here \texttt{pclass} is the cabin class,
\texttt{sex} is M or F, and the
passenger's \texttt{age} is in years.
The variables \texttt{sibsp} and \texttt{parch} 
count the number of siblings+spouses and 
parents+children aboard. 

\begin{figure}[!ht]
\centering
\includegraphics[width=.75\textwidth]
   {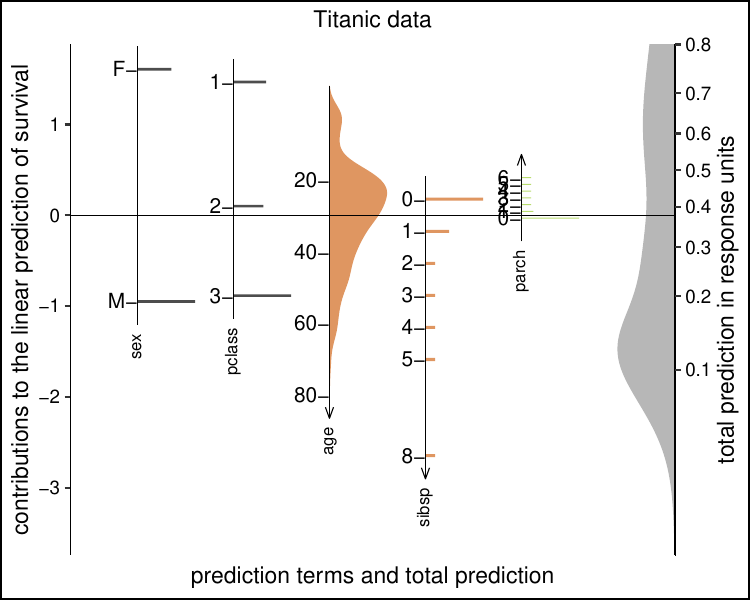}
\vspace{-2mm}
\caption{Predictions plot of the logistic 
regression on the Titanic data. The prediction
terms are again on the linear scale of the 
vertical axis on the left, but the labels of
the total prediction on the right are not
equally spaced since they indicate the
predicted probability of survival. Here
the continuous distributions are displayed by
estimated densities instead of histograms.}
\label{fig:titanic_predsplot}
\end{figure}

Figure~\ref{fig:titanic_predsplot} shows the
predictions plot. It plots the $f_j$ and the 
total linear prediction $\sum_j f_j$ as 
before, but the labels on the rightmost axis 
are on the scale of the predicted survival 
probability $g^{-1}(\sum_j f_j)$.
This time we opted to plot estimated
densities of \texttt{age} and the total 
prediction instead of their histograms.
We see at first glance that the gender of the
passenger is the most important variable, 
with predicted survival higher for females.
This catches the eye to a similar extent as 
regression trees on this dataset making 
their first split on this variable, see e.g. 
\cite{NNtree2022}. Next up is the cabin
class, followed by \texttt{age} which has a
down arrow, meaning that younger people had
a better chance at survival. The 
prediction terms \texttt{sex} and 
\texttt{age} reflect the motto `women and 
children first in the lifeboats,' but being 
a first class passenger increases the 
prediction too. Clearly variable 
\texttt{parch} has only a tiny effect.

It is hoped that the way the predictions plot
visually combines much information provides 
helpful insight in the output of a regression.
Figures~\ref{fig:topgear_predsplot} 
and~\ref{fig:titanic_predsplot} were obtained 
by the function \texttt{predsplot} in the 
\textsf{R} package \texttt{classmap} on
CRAN, using 
the single command \texttt{predsplot(fit)} 
where \texttt{fit} is an output object of 
the function \texttt{lm()} or \texttt{glm()}. 
The function \texttt{predsplot} figures 
out the rest. 
Of course there are many options: the 
maximal number of prediction terms to 
be shown, graphical parameters for colors and 
line widths, and so on. There is also 
an argument to specify whether to display 
histograms or estimated densities. For
densities there are arguments that input the 
desired bandwidth as well. Here we used the 
defaults of \texttt{stats::density()}.
The formula of the regression may
include logarithms of variables and
interaction terms, such as
$y \sim x_1 + \log(x_2) + x_3:x_4$
as illustrated in Section A of the 
Supplementary Material.

\section{Why is the prediction for my case
         so high (low)?}
\label{sec:casetoshow}

\begin{figure}[!ht]
\centering
\includegraphics[width=.85\textwidth]
   {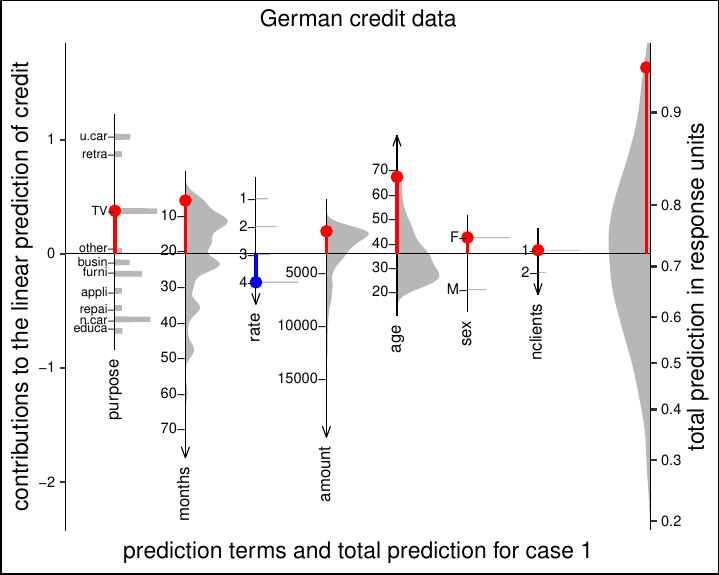}
\vspace{-2mm}
\caption{Predictions plot for case 1 of 
the German credit dataset. This loan
application's high predicted success 
probability on the right, over $0.9$ and 
shown in red, is explained by the mostly 
positive (red) values of its prediction
terms, with only a single negative (blue)
term. The distributions of the prediction
terms are not colored, to put the focus on 
the colors and sizes of the prediction 
terms of the case being displayed.}
\label{fig:germancredit_case_1}
\end{figure}

The predictions plot can also be used to
explain the prediction of a single case, which 
may be one of the cases the model was fitted on 
(`in-sample'), or a new case (`out-of-sample') 
for which we need a prediction.
We illustrate this with another benchmark, the
German credit dataset which is available from
\url{https://archive.ics.uci.edu/ml/datasets/statlog+(german+credit+data)} 
and from the \textsf{R} package \texttt{fairml} 
\citep{Scutari2023}. It contains
1000 loan applications. We predict the success 
probability of a loan by logistic regression on
the size of the loan (\texttt{amount}), its
duration (\texttt{months}), the interest 
\texttt{rate}, the intended \texttt{purpose} 
of the loan (categorical with 10 levels), the 
number of clients responsible for paying back
the loan (\texttt{nclients}), and the main 
client's \texttt{sex} and \texttt{age}.
Figure~\ref{fig:germancredit_case_1} shows
the predictions plot for the first case 
in the dataset.

In the predictions plot we see that the 
\texttt{purpose} of the loan has the largest 
effect, followed by the three financial 
variables. The latter have downward 
arrows, so a longer duration  
(\texttt{`months'}), a higher 
\texttt{rate}, and  a higher \texttt{amount} 
each decrease the predicted success probability. 
The variable \texttt{age} points upward, the 
model deems females more reliable than males, 
and being a one-client loan is considered 
positive. Now all the distributions are shown 
in grey, because the focus is on the prediction 
values. Those are red when the prediction value 
is above average, and blue if below average. 
From Figure~\ref{fig:germancredit_case_1} we
can see that it is about a woman in her 
sixties requesting a small
loan with short duration to buy a TV. Her 
prediction terms are all above average except
the one due to the high interest rate. Therefore
the total prediction on the rightmost axis is
very high, with estimated success probability 
over 90\%. It is also possible to add a 
profile curve to this plot, as illustrated in 
Section C of the Supplementary material.

\begin{figure}[!ht]
\centering
\includegraphics[width=.85\textwidth]
   {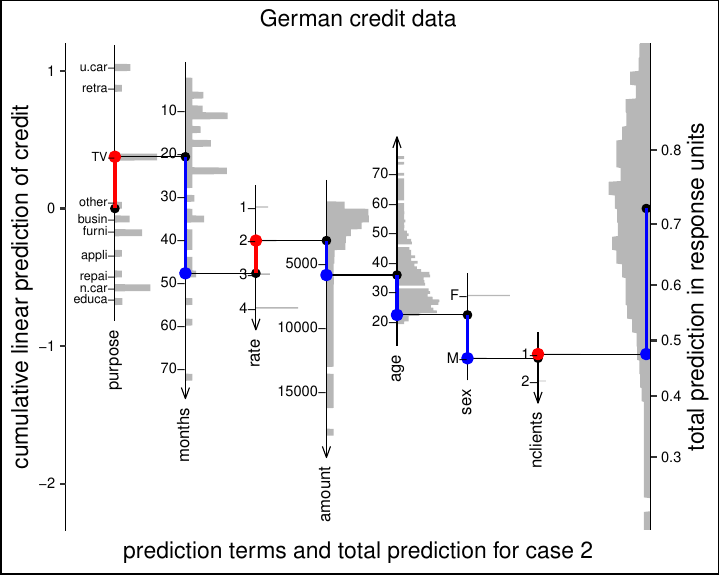}
\vspace{-2mm}
\caption{Predictions plot for case 2 of 
the German credit dataset. In this `staircase
style' the prediction terms are added up from
left to right, so after the last term we
arrive at the total linear prediction. This 
particular loan application has many 
negative (blue) prediction terms, yielding 
a negative total linear prediction that 
translates to a relatively low success 
probability.}
\label{fig:germancredit_case_2}
\end{figure}

Figure~\ref{fig:germancredit_case_1} suggests 
that the individual prediction terms 
are added, but does not actually show that. A 
variation of the plot that does visualize the 
addition is available as the option 
\texttt{staircase = TRUE}, illustrated
in Figure~\ref{fig:germancredit_case_2} for 
case 2 of the dataset. In this plot the 
prediction terms are added by shifting each 
prediction term up or down so its
average aligns with the cumulative prediction 
to its left. We see that case 2 is a young man
requesting a larger loan with longer duration,
and his prediction of credit worthiness comes 
out rather low. If a loan is refused based on 
such a prediction, a graph like 
Figure~\ref{fig:germancredit_case_1}
or Figure~\ref{fig:germancredit_case_2}
facilitates explaining why to the client.

The choice between the basic style in 
Figure~\ref{fig:germancredit_case_1} and the
staircase style in 
Figure~\ref{fig:germancredit_case_2} is a 
matter of taste. Whereas the latter shows 
the addition explicitly, the plot also looks 
busier, and its appearance depends more
strongly on the order of the individual
prediction terms. The choice is left 
to the user. More examples of both styles are 
shown in the Supplementary Material.

While it can be very useful or even mandatory 
to be able to explain a prediction as in 
Figure~\ref{fig:germancredit_case_2},
making the entire prediction 
equation~\eqref{eq:predglm} public
may have the undesired effect that people 
can game the system to obtain their  
favored outcome. After seeing  
Figure~\ref{fig:germancredit_case_2}, the
client might be tempted to increase his
chances of getting a loan by indicating the
intended purpose is to buy a used car because
the \texttt{u.car} level is at the top, 
selecting a smaller number of months, and 
asking his mother to fill in the application 
form to get more favorable values 
of \texttt{age} and \texttt{sex}, while 
keeping the other input variables 
unchanged. Running \texttt{predsplot} 
on this new set of input variables 
yields the figure in Section C of the 
Supplementary Material, in which the 
predicted probability becomes 89\% 
instead of 48\%\,.

\section{Correlations between prediction terms} 
\label{sec:predscor}
      
It is well-known that the coefficient of an 
input variable in multiple regression 
can have the opposite sign of its slope in a 
simple regression with that input 
variable alone. This often happens when that
input variable is correlated with other 
input variables in the multiple model. 
Therefore it is useful to look at the 
correlations between input variables.
One often visualizes these correlations
by a heatmap, in which correlations of zero
are shown in white, correlations close to 1
in dark red, and correlations close to -1 
in dark blue, with a gradual change in 
between the extremes. 

\begin{figure}[!ht]
\centering
\includegraphics[width=.98\textwidth]
   {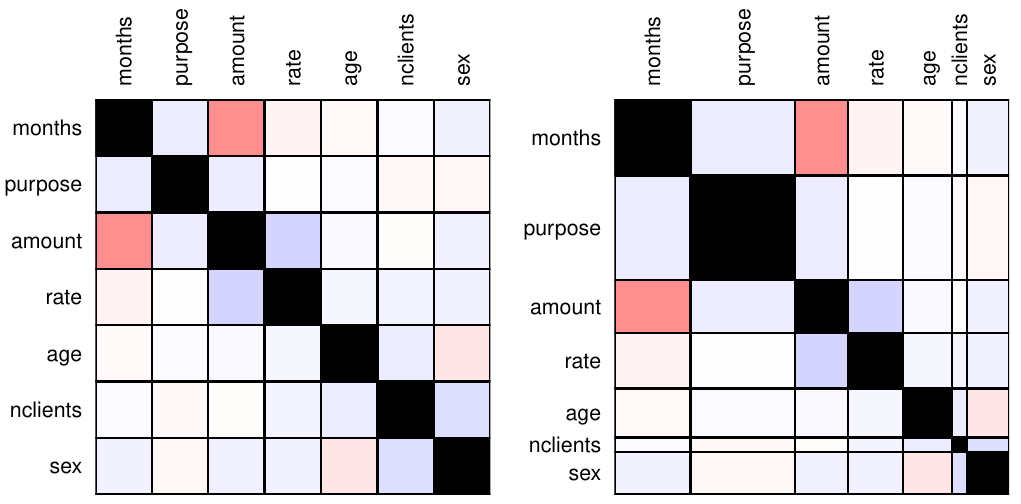}   
\caption{Correlations between the
prediction terms of the German credit data. 
The black squares on the diagonal have
correlation 1. A red cell indicates a
positive correlation, with color intensity
proportional to the correlation value.
A negative correlation is shown in blue,
and zero correlation becomes white.
The left panel is a customary correlation
plot in which all cells have the same size. 
The right panel is the proposed version, 
where the side of a diagonal square is 
proportional to the standard deviation of 
its prediction term. The area of each 
off-diagonal cell is then proportional to 
the covariance of its prediction terms.}
\label{fig:predscor_credit_1}
\end{figure}

We propose to enhance such a display 
in two ways. First, by switching from 
input variables to 
prediction terms, so that also
categorical input variables can be 
included. The left panel of 
Figure~\ref{fig:predscor_credit_1}
shows such a display for the German
credit data. For a numerical 
input variable, the sign of its slope
coefficient matters. For a categorical
input variable, its 
prediction term is numerical so
its correlation with other 
prediction terms exists.

Note that this does not yet take the effect 
magnitude of the prediction terms into 
account. Our second modification is to plot 
the diagonal cells with sides that are 
proportional to the standard deviation of 
their prediction term. In this way 
the area of each diagonal cell is 
proportional to the variance of its 
prediction term. The right panel 
of Figure~\ref{fig:predscor_credit_1}
shows the result. We immediately 
see that the variable \texttt{purpose}
has a big effect, whereas that of
\texttt{nclients} is the smallest.
Interestingly, such a correlation display
with varying cell sizes did not appear
to exist yet in the literature or on the
internet, nor heatmaps of that type.

The off-diagonal part of the display
consists of rectangles, which also have
a meaningful interpretation. The area of
such a rectangle is the product of the
standard deviations of its 
prediction terms, and when we 
multiply that with the value 
of the correlation (represented by its 
color) we obtain the covariance between 
the prediction terms. So a dark 
color is more important in a larger 
rectangle (e.g. formed by \texttt{months} 
and \texttt{amount}) than in a smaller 
rectangle (e.g. formed by \texttt{rate}
and \texttt{amount}). 
The proposed display thus depicts the 
covariance matrix between the 
prediction terms, which is relevant 
in a regression. Such a display would 
not be meaningful for a covariance matrix 
between variables in different units, 
but here all prediction terms 
are in the same units (those of the total 
linear prediction).

The argument of the \textsf{R} function 
\texttt{predscor()} making this 
display is again a fit produced by
\texttt{lm()} or \texttt{glm()}.
There are various
options, such as sorting the
prediction terms by decreasing 
standard deviation, plotting the absolute 
values of the correlations, or making the 
area of the diagonal cells proportional 
to the standard deviation of the 
prediction terms rather than their 
variance.

To illustrate what can happen when two 
input variables are highly correlated, 
we carry out a small experiment. We replace 
the input variables \texttt{months} 
and \texttt{nclients} by their sum 
\texttt{x1 = months + nclients} 
and their difference 
\texttt{x2 = months - nclients}.
Of course this is an artificial example, 
but it does yield two 
input variables with a correlation 
over $0.99$, and the same total prediction.

\begin{figure}[!ht]
\centering
\includegraphics[width=.85\textwidth]
   {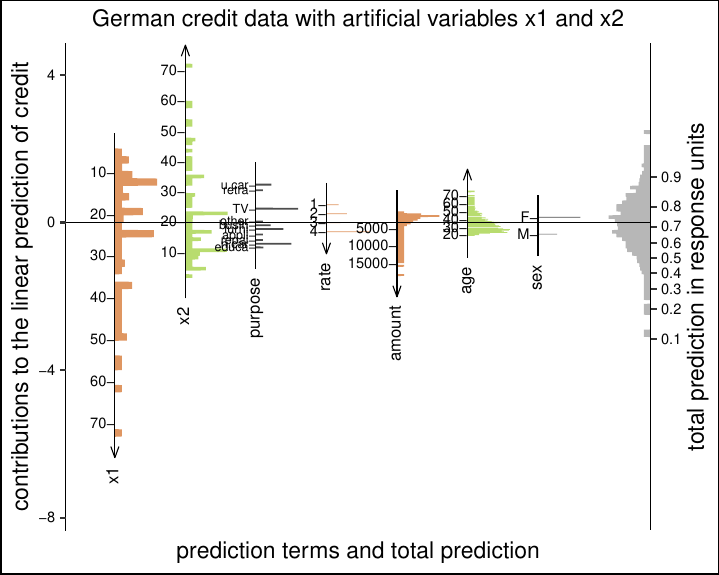}
\vspace{-2mm}
\caption{Illustration of the issue
of highly correlated input variables, that
is, near-multicollinearity.
The input variables \texttt{months} and 
\texttt{nclients} of the German credit data 
are replaced by two artificial input 
variables \texttt{x1} and \texttt{x2} that
have a large positive correlation between 
them. In this predictions plot their
prediction terms point in opposite 
directions, indicating that their slope 
coefficients have opposite signs. These 
prediction terms are more spread out than 
the total linear prediction.}
\label{fig:predsplot_credit_artif}
\end{figure}

\begin{figure}[!ht]
\centering
\includegraphics[width=.49\textwidth]
   {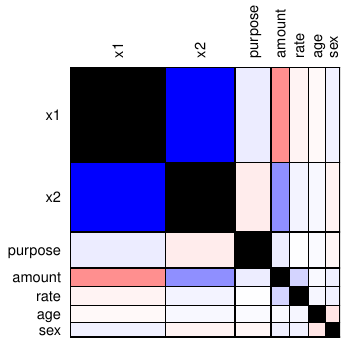}
\caption{Correlations between the
prediction terms in
Figure~\ref{fig:predsplot_credit_artif}.
The large diagonal squares labeled 
\texttt{x1} and \texttt{x2} reflect their 
high variance, and the dark blue cell 
between them indicates their large 
negative correlation.}
\label{fig:predscor_credit_2}
\end{figure}

Figure~\ref{fig:predsplot_credit_artif} 
shows the resulting predictions plot. 
It looks rather suspicious, because
the artificial variables \texttt{x1} and
\texttt{x2} are more spread out than the
total prediction on the right, which is
the sum of these and other terms. 
Moreover, the predictions of \texttt{x1} 
and \texttt{x2} look like mirror images. 
They partly cancel each other. 
The correlation display in 
Figure~\ref{fig:predscor_credit_2} also
looks unusual, with the big dark blue
rectangles between \texttt{x1} and 
\texttt{x2}. The correlation between 
\texttt{x1} and \texttt{x2} that was close
to 1 has flipped sign between their
prediction terms, again reflecting 
the canceling effect. 
In general, prediction terms are 
easiest to interpret when they only have 
small correlations with each other. 

\section{Related methods}
\label{sec:related}

The predictions plot is a member of the 
extensive family of \textit{parallel 
coordinate plots} \citep{Inselberg2009}.
A related plot is the nomogram 
for binary logistic regression proposed
by~\cite{Zlotnik2015}. It is also a parallel 
plot of prediction terms, but its purpose is
computation rather than interpretation.
The user needs to read off a score on each
variable axis, add up the scores manually, 
and then look up the sum on another axis 
to obtain the resulting approximate 
probability. 
No distributions are visualized, and the 
predictions are not centered but 
instead their minima are aligned. 
This makes the nomogram hard to interpret, 
but interpretation was not its objective.

\textit{Effect plots} \citep{Fox2003} describe 
the effect of varying a single input variable 
on the total prediction. During this computation
the other input variables are kept fixed at
typical values, such as their average or 
median. A variation on this idea is to use the
values taken by the remaining input variables
in the actual dataset with $n$ cases, and then 
to average the $n$ results. Either way one 
obtains a plot of the total prediction as a 
function of the input variable being studied.
The effect need not be linear, as the effect
plot can be drawn for generalized linear 
models. The purpose of effect plots is 
similar to ours, to gain insight in the
prediction, but the fact that it focuses on 
one input variable at a time makes it quite
different from the approach presented here.

A {\it coefficient plot}, described  
for instance by \cite{Arel2022}, does look 
at several input variables at the same time. 
But it does not plot the distributions of 
the input variables or the prediction. 
Instead it plots the regression coefficients 
associated with the input variables, that
need to be numerical, as well as their
confidence intervals. The coefficients are
plotted as points, and the confidence intervals
as line segments. The coefficients can be 
made comparable to each other if the input 
variables are first standardized, say to 
zero mean and unit standard deviation.
The coefficient plot describes the whole 
model, but in a way very different from the 
predictions plot.

A \textit{forest plot}, see for instance 
\cite{Chang2022}, looks rather similar to 
a coefficient plot. But it displays the
outcomes of different studies in a 
meta-analysis, with their confidence intervals.
The predictions plot instead shows the 
distributions of the prediction terms in a 
single dataset.

\section{Discussion}
\label{sec:conclusions}

There are many other ways to measure the 
contribution of input variables. 
\cite{Verdinelli2024} reviewed several, and 
found that there can be no `neutral' way to 
do so, and that there is currently no 
`overall best' way. The displays in this note 
may seem oversimplified but that makes them 
easy to interpret. Their main benefit is that 
they help explainability of the prediction 
rule as it is given, without depending on how 
it was derived.
It would be possible to extend the displays 
to other settings, but that falls outside 
the scope of this note.\\

\noindent{\bf Software Availability.} The
functions \texttt{predsplot()} and
\texttt{predscor()} have been added to the
\textsf{R} package \texttt{classmap}
\citep{classmap} on CRAN, together with a 
vignette that reproduces all the figures 
in the paper and the supplementary text.\\

\noindent{\bf Acknowledgments.} Thanks go to
Mia Hubert, Jakob Raymaekers, and the 
reviewers for helpful suggestions that 
improved the presentation.\\

\noindent{\bf Disclosure Statement.} The 
author reports there are no competing 
interests to declare.

\clearpage

\pagenumbering{arabic}

\appendix

\setcounter{figure}{7} 
   
\begin{center}

\LARGE{\bf Supplementary Text}\\
\end{center}

\section{More figures of the Top Gear data}
\label{suppmat:cars}

We start with the example in the
introduction, the fit
$$\widehat{hp} = 2.466*\mbox{topspeed}
  -13.13*\mbox{length}
  + 0.063*\mbox{displacement} -206.9$$
that was obtained from the standard regression
\begin{verbatim}
> fit = lm(hp ~ topspeed + length + displ, data = cars)
>
> summary(fit)
# Coefficients:
#               Estimate Std. Error t value Pr(>|t|)    
# (Intercept) -2.069e+02  2.861e+01  -7.232 4.78e-12 ***
# topspeed     2.466e+00  1.373e-01  17.963  < 2e-16 ***
# length      -1.313e+01  6.187e+00  -2.122   0.0347 *  
# displ        6.255e-02  2.808e-03  22.275  < 2e-16 ***
\end{verbatim}
It is a simple example because all the input
variables are numeric. The question was, which of 
these has the biggest effect on the prediction. 
Using \texttt{predsplot()} we can easily answer this:
\begin{verbatim}
> library("classmap")
> predsplot(fit, main = "Top Gear data")
#         prediction term   stdev up/down
#                topspeed  68.380      up
#                  length   5.817    down
#                   displ  91.790      up
#  Total prediction of hp 149.200
\end{verbatim}
We see that the prediction term from \texttt{displ} 
has the largest standard deviation, followed by
\texttt{topspeed}, whereas \texttt{length} only
has a tiny effect.
The resulting predictions plot visualizes this:
\clearpage
\begin{figure}[!ht]
\centering
\includegraphics[width=.8\textwidth]
   {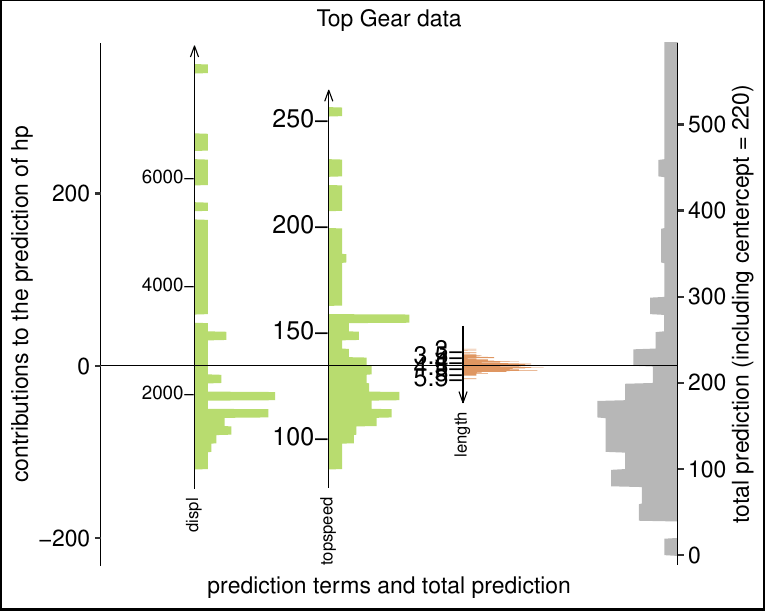}   
\end{figure}

We clearly see that the spread of the prediction
term of \texttt{length} is tiny.
This is confirmed by the new display of the
correlations and standard deviations of the 
prediction terms:
\begin{verbatim}
> predscor(fit, sort.by.stdev = FALSE)
\end{verbatim}
\begin{figure}[!ht]
\centering
\includegraphics[width=.43\textwidth]
   {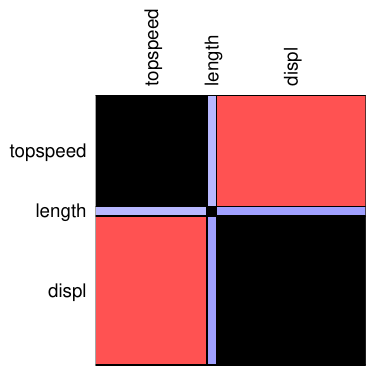}   
\end{figure}

\newpage
For Figure~\ref{fig:topgear_predsplot} in 
the main text we first ran a standard regression by

\begin{verbatim}
> fit = lm(GPM ~ accel + drive + weight + fuel, data = cars)
> summary(fit)
\end{verbatim}
yielding, among other things:

\begin{verbatim}
Coefficients:
              Estimate Std. Error t value Pr(>|t|)    
(Intercept)  1.681e-02  4.884e-03   3.443 0.000677 ***
accel       -1.356e-03  2.459e-04  -5.516 8.81e-08 ***
driveFront  -2.927e-03  1.619e-03  -1.808 0.071853 .  
driveRear   -2.745e-04  1.577e-03  -0.174 0.861992    
weight       1.131e-05  1.841e-06   6.143 3.25e-09 ***
fuelPetrol   8.427e-03  1.246e-03   6.761 9.98e-11 ***
\end{verbatim}
The function \texttt{lm()} has encoded the 
categorical input variable 
\texttt{drive} with 3 levels by two
binary dummy variables \texttt{driveFront} and
\texttt{driveRear}, and the categorical variable 
\texttt{fuel} by the single binary dummy 
\texttt{fuelPetrol}.
From this summary we cannot derive the
relative effect of the four input variables
\texttt{accel}, \texttt{drive}, \texttt{weight}
and \texttt{fuel}.\\ 

For the numerical input 
variables \texttt{accel} and \texttt{weight}
there is a known remedy, which is to divide 
them by their standard deviation and then rerun 
the fit:
\begin{verbatim}
> cars$st.accel  = cars$accel/sd(cars$accel)
> cars$st.weight = cars$weight/sd(cars$weight)
> st.fit = lm(GPM ~ st.accel + drive + st.weight + fuel, data = cars)
> summary(st.fit)
\end{verbatim}
yielding

\newpage
\begin{verbatim}
Coefficients:
              Estimate Std. Error t value Pr(>|t|)    
(Intercept)  0.0168141  0.0048841   3.443 0.000677 ***
st.accel    -0.0043289  0.0007848  -5.516 8.81e-08 ***
driveFront  -0.0029267  0.0016189  -1.808 0.071853 .  
driveRear   -0.0002745  0.0015774  -0.174 0.861992    
st.weight    0.0044897  0.0007308   6.143 3.25e-09 ***
fuelPetrol   0.0084270  0.0012465   6.761 9.98e-11 ***
\end{verbatim}
This changes the coefficients of acceleration 
and weight, all other coefficients remaining
the same. From the absolute values of the 
new coefficients we see
that the effect sizes of acceleration and weight
are similar to each other, with that of weight 
being a bit higher. But we cannot do the same
with a categorical variable like \texttt{drive} 
that is encoded by two dummies. Should we 
standardize both dummies, and even if we did, 
how can we combine their effects?\\

This is where the predictions plot comes in
handy. We just run
\begin{verbatim}
> library("classmap")
> main = "Top Gear Data"
> predsplot(fit, main = main)
\end{verbatim}
yielding Figure~\ref{fig:topgear_predsplot}, 
and on the terminal we get the standard 
deviations of the prediction terms:
\begin{verbatim}
#          prediction term    stdev up/down
#                    accel 0.004329    down
#                    drive 0.001400        
#                   weight 0.004490      up
#                     fuel 0.004104        
#  Total prediction of GPM 0.009783
\end{verbatim}
We see that the effect of the input variable 
\texttt{drive} is the smallest of the four.\\

The correlations between the four prediction
terms are visualized by the line
\begin{verbatim}
> predscor(fit)
\end{verbatim}

\clearpage
\begin{figure}[!ht]
\centering
\includegraphics[width=.45\textwidth]
   {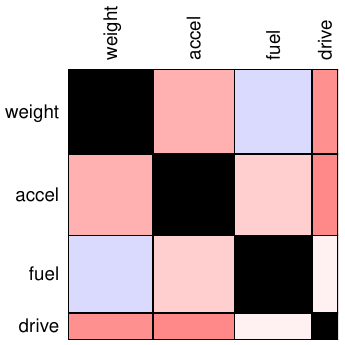}   
\end{figure}
\noindent where we again see that the black 
square of \texttt{drive} is the smallest.\\

We can also make a predictions plot for a
single case, in order to explain its prediction.
To illustrate this we choose two rather extreme 
cars, the Bentley Continental and the Kia Rio:
\begin{verbatim}
> car = "Bentley Continental"
> predsplot(fit, main = main, casetoshow = car, displaytype = "density",
                 xlab = paste0("prediction for ", car))
> car = "Kia Rio"
> predsplot(fit, main = main, casetoshow = car, staircase = TRUE,
               xlab = paste0("prediction for ", car)) 
               
\end{verbatim}

In the first plot we see that the Bentley
gets a very high predicted gallons per mile 
(GPM). The prediction plot (this time with 
densities) explains its gas guzzling 
behavior by the fact that it is heavy (around 
2300 kg as seen in the plot), accelerates 
from standstill to 60 miles per hour in 
under 5 seconds, runs on petrol, and has 
all-wheel drive (4WD).

The Kia Rio in the next plot finds 
itself in the opposite situation, as it 
is predicted to require very little fuel. 
The plot (this time in staircase style) 
explains this by the car's low weight
(around 1200 kg in the plot), slow 
acceleration, diesel engine, and 
front wheel drive.
\clearpage

\begin{figure}[hb]
\centering
\includegraphics[width=.79\textwidth]
   {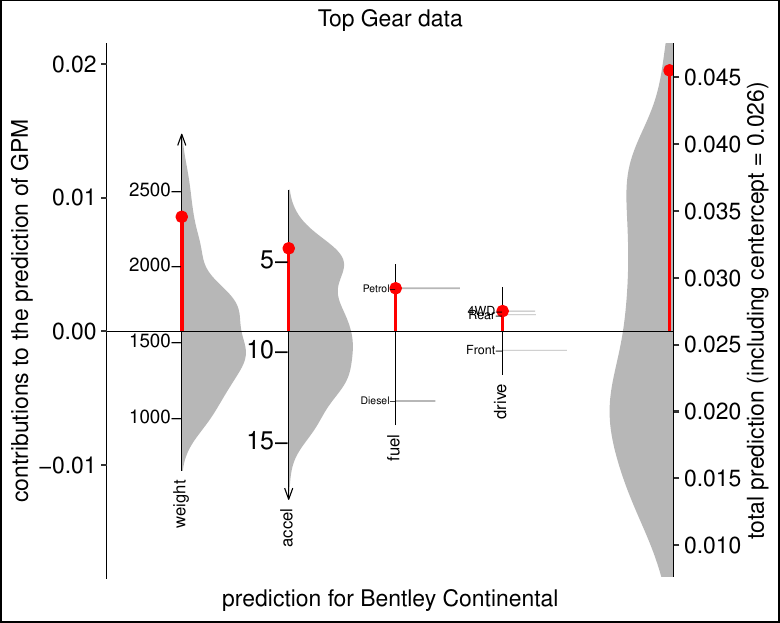}\\

\vspace{4mm}

\includegraphics[width=.79\textwidth]
  {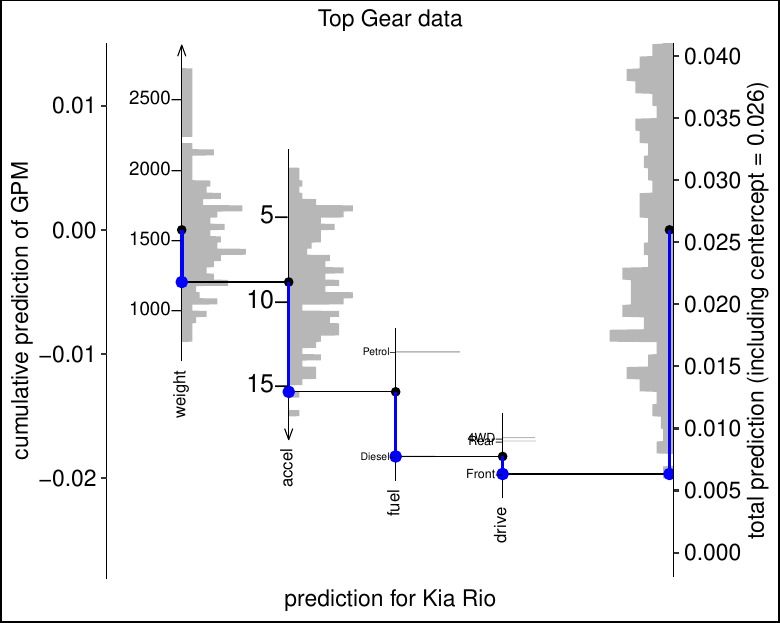}
\end{figure}
\clearpage

To illustrate that the code can handle 
transformations and interactions in the 
formula we run

\noindent $> \texttt{fit = lm(1/MPG } \sim 
\texttt{ accel + log(weight) + 
accel:torque, data = cars)}$

\noindent This gets picked up automatically in 
the displays below. Interactions between 
categorical input variables, or between
a numerical and a categorical variable, are 
visualized in the same way.

\begin{figure}[!ht]
\centering
\includegraphics[width=.45\textwidth]
   {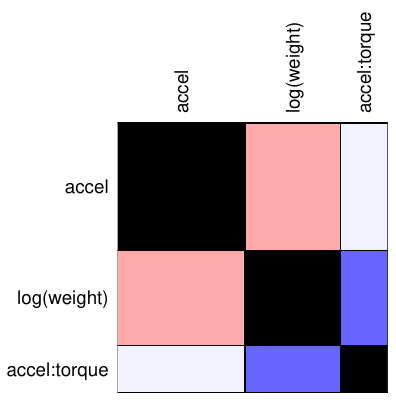}  
   
\vspace{6mm}

\includegraphics[width=.76\textwidth]
   {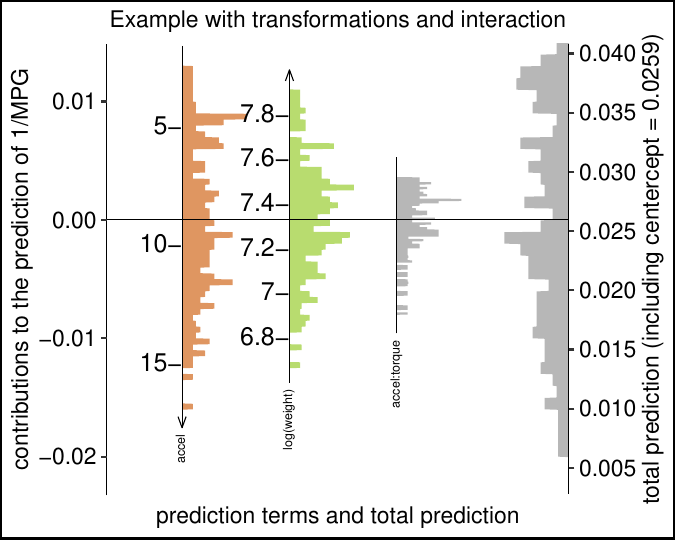}
\end{figure}

The code can also deal with logical
and character variables. Let us run

\noindent $> \texttt{fit = lm(1/MPG } \sim 
\texttt{accel + log(weight) + 
accel:torque + alarm + navig)}$\\
where \texttt{alarm} (alarm system) is 
TRUE/FALSE, and \texttt{navig} (satellite
navigation) is ``y"/``n".

\begin{figure}[!ht]
\centering
\includegraphics[width=.48\textwidth]
   {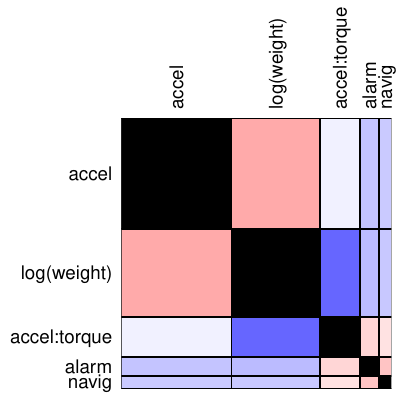}
\vspace{6mm}

\includegraphics[width=.79\textwidth]
   {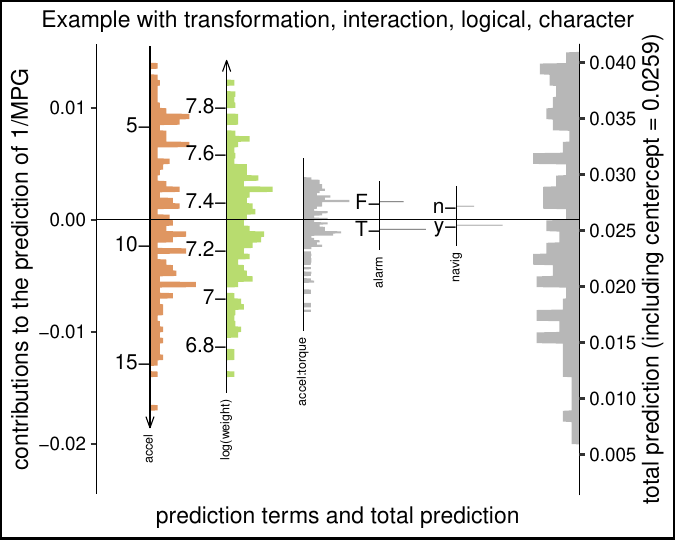}
\end{figure}

\clearpage
\section{More figures of the Titanic data}
\label{suppmat:titanic}

We now draw the correlation display of the Titanic
logistic regression, by the lines
\begin{verbatim}
> fit <- glm(y ~ sex + age + sibsp + parch + pclass,family=binomial,data=titanic)
> predscor(fit, sort.by.stdev = FALSE)
\end{verbatim}

\begin{figure}[!ht]
\centering
\includegraphics[width=.49\textwidth]
   {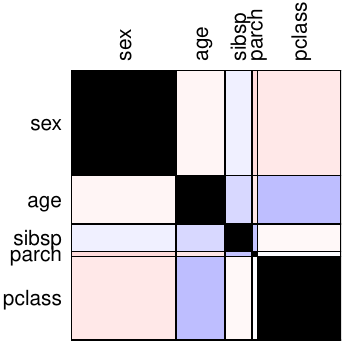}
\end{figure}
\noindent It clearly indicates
that the passenger's sex has a large effect on
the prediction, followed by passenger class and
age. On the other hand, the effect of 
\texttt{parch} is minuscule. The off-diagonal
colors are rather light, indicating smaller
correlations between the Titanic 
prediction terms than between those of 
the Top Gear data.

The predictions plots for passengers 1 and 2 
on the next page are made by
\begin{verbatim}
> predsplot(fit, main = main, casetoshow=1)
> predsplot(fit, main = main, casetoshow=2, staircase = TRUE)
\end{verbatim}

In the first plot we see that the
predicted survival probability of
passenger 1 is low. Most of the 
prediction terms
are negative, mainly because the passenger 
was male and traveled in third class. The 
only positive prediction term comes 
from his young age. 

Passenger 2 is a woman traveling
in first class, so in spite of the slightly
negative age-based prediction term 
her total prediction is relatively high.

\begin{figure}[!ht]
\centering
\includegraphics[width=.77\textwidth]
  {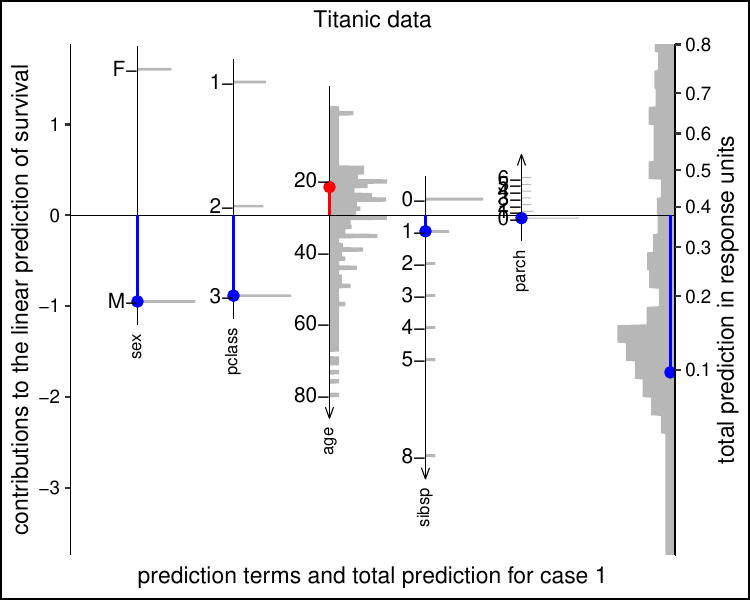}

\vspace{8mm}

\includegraphics[width=.77\textwidth]
  {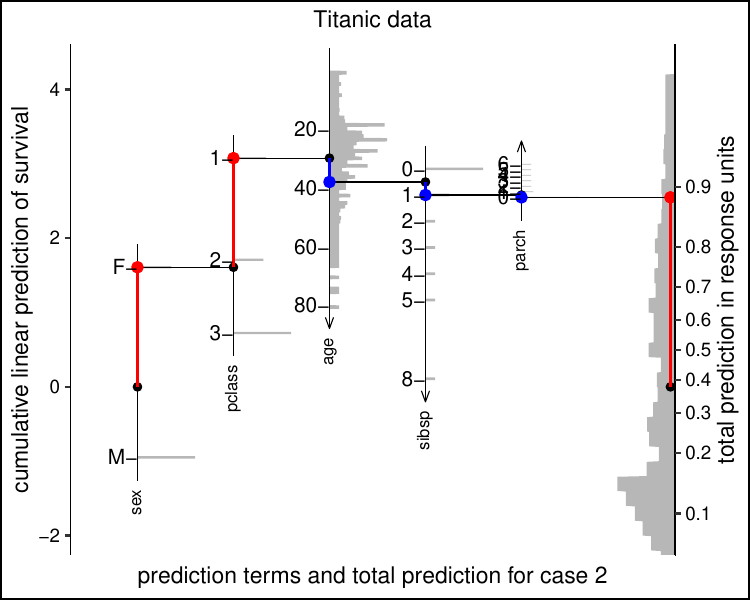}
\end{figure}

\clearpage
\section{More figures of the German credit data}
\label{suppmat:credit}

Here is the overall predictions plot of the
German credit data:

\vspace{2mm}
\begin{figure}[!ht]
\centering
\includegraphics[width=1.0\textwidth]
  {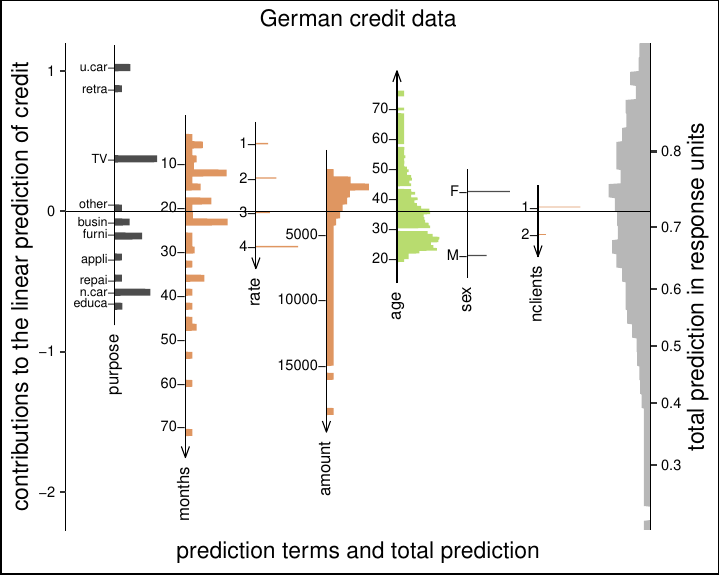}
\end{figure}

\vspace{2mm}
Next we look at a hypothetical new case, that is 
not in the original (training) dataset but was
motivated at the end of 
Section~\ref{sec:casetoshow}.
The new case is given as follows:

\begin{verbatim}
> newc = list("u.car", 36, 2, 6000, 55, "F", 1)
> names(newc) = c("purpose", "months", "rate", "amount", "age", "sex", "nclients")
> predsplot(fit, main = main, casetoshow = newc, staircase = TRUE)
\end{verbatim}

\noindent This yields the plot on the next page:

\clearpage
\begin{figure}[!ht]
\centering
\includegraphics[width=.82\textwidth]
  {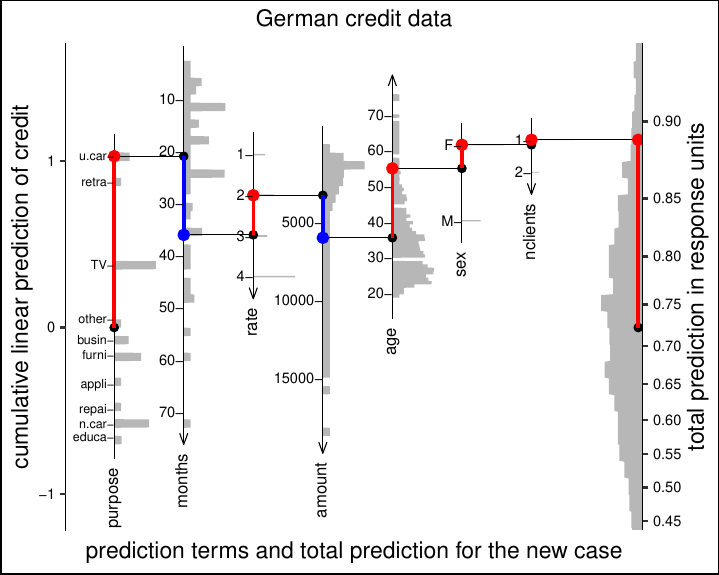}
\end{figure}

\noindent In the output on the terminal we see that 
the new predicted probability is about $89\%$\,:

\begin{verbatim}
                              prediction term    value
                                       months -0.47190
                                      purpose +1.02816
                                       amount -0.25499
                                         rate +0.23763
                                          age +0.41640
                                     nclients +0.03030
                                          sex +0.14143
                                          SUM +1.12701
                                   centercept  0.95998
            Total linear prediction of credit  2.08699
 Total prediction of credit in response units  0.88963
\end{verbatim}

\newpage
Note that non-staircase prediction plots for 
a single case, like 
Figure~\ref{fig:germancredit_case_1}, can also 
be shown with a profile by the argument
\texttt{profile = TRUE} :
\begin{verbatim}
> predsplot(fit, main = main, casetoshow = 1, 
            displaytype = "density", profile = TRUE)
\end{verbatim}

\vspace{6mm}
\begin{figure}[!ht]
\centering
\includegraphics[width=.82\textwidth]
  {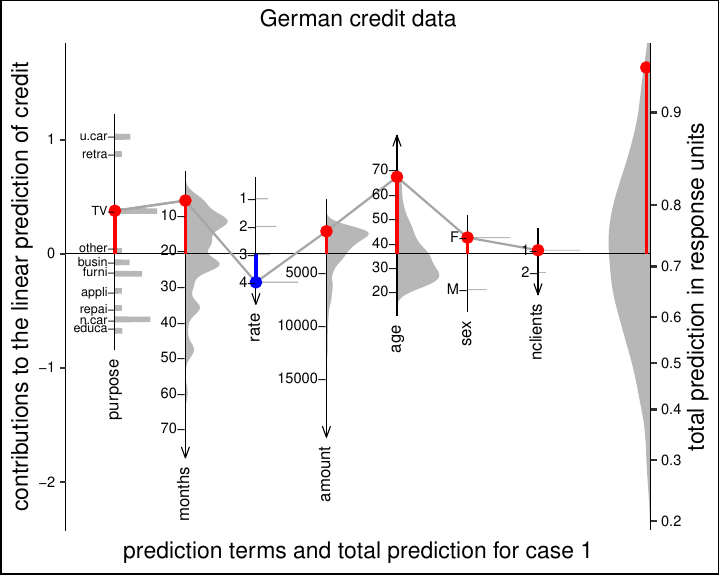}
\end{figure}

\vspace{5mm}
The profile is the faint grey broken line that connects
the prediction terms of this case. It makes it easier
to compare plots of this type made for different cases.

\end{document}